\documentclass[prd,preprint
               showpacs,                                       
               preprintnumbers,aps,epsfig,%a4paper,
               nofootinbib]{revtex4}
\usepackage{graphicx}
\usepackage{amssymb,latexsym}
\usepackage{amsmath}
\newcommand{\beq}{\begin{equation}}
\newcommand{\eeq}{\end{equation}}
\newcommand{\beqa}{\begin{eqnarray}}
\newcommand{\eeqa}{\end{eqnarray}}
\newcommand{\mpl}{M_{Pl}}
\newcommand{\lmk}{\left(}
\newcommand{\rmk}{\right)}
\newcommand{\lkk}{\left[}
\newcommand{\rkk}{\right]}

\newcommand{\calF}{{\cal F}}

\newcommand{\vecs}[1]{\mbox{\boldmath\tiny${#1}$}}
\newcommand{\vect}[1]{\mbox{\boldmath${#1}$}}
\newcommand{\mg}{M_{\rm Pl}}
\newcommand{\piv}{\ast}

\usepackage{color}
\begin{document}

\preprint{RESCEU-44/14}

\date{\today}

\title{%
 Prospects of determination of reheating temperature
after inflation by DECIGO %in the light of recent BICEP2 result
}
\author{%
Sachiko Kuroyanagi$^1$, Kazunori Nakayama$^2$, 
and Jun'ichi Yokoyama$^{3,4}$
}
\address{%
$^1$Department of Physics, Tokyo University of Science,
Kagrazaka, Japan\\
$^2$Department of Physics, Graduate School of Science,  
The University of Tokyo, 
Tokyo 113-0033, Japan\\
$^3$Research Center for the Early Universe (RESCEU),
Graduate School of Science, 
The University of Tokyo,
Tokyo 113-0033, Japan\\
$^4$Kavli Institute for the Physics and Mathematics of the Universe
 (Kavli IPMU), WPI, TODIAS, The University of Tokyo, Kashiwa, Japan\\
}

%%%%%%%%%%%%%%%%%%%%%%%%%%%%%%%%%%%%%%%%%%%%%%%%%%%%%%%%%%%%%%%%%%%%%%%%%%%
\begin{abstract}
 If the tensor-to-scalar ratio $r$ of cosmological perturbations takes a
 large value $r\sim 0.1$, which may be inferred by recent BICEP2 result,
 we can hope to determine thermal
  history, in particular, the reheating temperature, $T_R$, after
  inflation by space-based laser interferometers.  It is shown that
     upgraded and upshifted versions of DECIGO may be able to determine 
$T_R$ if it lies in the range $6\times 10^6< T_R < 5\times 10^7$GeV
and $3\times 10^7<T_R<2\times 10^8$GeV, respectively.  Although these
 ranges include predictions of some currently plausible inflation
 models, since each specification can probe $T_R$ of at most a decade
 range, we should determine the specifications of DECIGO with full
 account of constraints on inflation models to be obtained by
 near-future observations of temperature anisotropy and B-model
 polarization of the cosmic microwave background radiation.
\end{abstract}
%%%%%%%%%%%%%%%%%%%%%%%%%%%%%%%%%%%%%%%%%%%%%%%%%%%%%%%%%%%%%%%%%%%%%%%%%%%


\maketitle

%\vskip 1cm
%\baselineskip 1cm
%\newpage

%%%%%%%%%%%%%%%%%%%%%%%%%%%%%%%%%%%%%%%%%%%%%%%%%%%%%%%%%%%%%%%%%%%%%%%%%%%

After more than three decades from its original 
proposal \cite{oriinf2,oriinf1}, the
inflationary cosmology is now confronting and passing
a number of observational tests.
Among its generic predictions, the spatial flatness and generation of almost
scale-invariant spectrum of curvature perturbations \cite{yuragi}
were first confirmed by precise measurements of the cosmic microwave
background radiation (CMB) by WMAP \cite{WMAP,WMAP9}
and are now being updated by Planck \cite{Ade:2013zuv,Ade:2013uln}. 
 The Gaussian nature of these 
fluctuations has also been  further confirmed recently 
by Planck 
\cite{Ade:2013ydc}.  

In March 2014, the BICEP2 collaboration \cite{bicep} reported detection of
the B-mode polarization of CMB over a fairly wide range of angular
multipoles  from $\ell \simeq 40$ to  350.  The higher multipole
range can be explained by gravitational lensing while the smaller
multipoles are interpreted as owing to the long-wave gravitational waves
of primordial origin, most likely from   
the tensor
perturbations generated quantum mechanically during inflationary
expansion stage in the early Universe \cite{staro}.
If what they measured had not been contaminated by foregrounds,
it would correspond to  the amplitude of the
tensor-to-scalar ratio as $r=0.20^{+0.07}_{-0.05}$ \cite{bicep}.
Note that  $r$ is related with the energy scale
of inflation as $V=(3.2\times 10^{16}{\rm GeV})^4r$.

This value, on the other hand,
 is larger than that expected by the constraints imposed by WMAP
\cite{WMAP9} and
Planck \cite{Ade:2013uln} in terms of temperature anisotropy and E-mode
 polarization,  because they reported 95\% upper bounds on $r$ as 
$r<0.13$ and 0.11, respectively, and that the 
likelihood contours in $(n_s,r)$ plane preferred the tensor-to-scalar
ratio significantly smaller than 0.1.  As a result, models predicting
tiny values of $r$ such as $r\sim 10^{-3}$ had been
investigated extensively including the curvature square inflation 
\cite{oriinf1}
and the original Higgs inflation model \cite{CervantesCota:1995tz},
which occupy the central region of the likelihood contours.
These models would be  ruled out if large tensor
 perturbation would be observationally established
\footnote{
Even in such a case the Higgs field
in the Standard Model could  be an inflaton, because
newer Higgs inflation models such as Higgs G-inflation
\cite{Kamada:2010qe} or running kinetic inflation \cite{Nakayama:2010kt}
could work well to accommodate large enough $r$
as summarized in \cite{Kamada:2012se}.}.

After the original announcement of BICEP2, several analyses 
of the effects of dust contamination
have been done, and it has been pointed out 
that they may be so large that the observed B-mode polarization may be
entirely due to the dust foreground \cite{Mortonson:2014bja}
 and we  only have an upper bound
on $r$.

In any event, the BICEP2 observation has reminded us the lesson that
the truth may not lie in the center of the likelihood contour and we
should remain open-minded until the final result is established.
Hence here we consider the case with $r$ close to its observational
upper bound $r\sim 0.1$.  The most plausible feature of a relatively
large value of $r$ is that direct observation of tensor perturbations
becomes more feasible by future space-based laser interferometers such
as DECIGO \cite{decigo}, which also allow us to extract useful
information on the thermal history after inflation \cite{Seto:2003kc}.
For example, information on reheating is imprinted in the
gravitational wave spectrum in the frequencies corresponding to the
energy scale of reheating.  Thus, the targeting frequency of the
experiment is a key for determining reheating temperature and would be
better to be adjusted once we obtain a hint about reheating from
either cosmology or particle physics.  In this paper, we discuss the
range of the reheating temperature which can be determined by DECIGO
and its upgraded and ultimate versions based on the updated
sensitivity curves than that used in \cite{Kuroyanagi:2011fy}.  We
also consider a specification whose sensitivity is shifted to higher
frequencies and see how the range of reheating temperature changes.

Specifically, we first consider the chaotic inflation model \cite{chaoinf}
driven by a massive scalar field for which  sensible particle physics
models exist \cite{Murayama:1992ua}.  This model predicts
tensor-to-scalar ratio $r= 0.13\sim 0.16$ at $N=60\sim 50$ $e$-folds
before the end of inflation, and so fits the lower tail of the BICEP2
result well, which may also be allowed by WMAP and Planck observations,
if not preferred.  Just in case a significant fraction of the BICEP2
result is due to the dust foregrounds, we also consider a natural
inflation model with the symmetry breaking scale $v=7\mpl$, which
predicts $r= 0.07 \sim 0.09$,
for comparison \cite{Freese:1990rb}.  Here $\mpl$ is the reduced Planck
scale. 

%Better agreement between BICEP2 and Planck/WMAP may be realized if we
%incorporate running spectral index \cite{bicep}.  Among inflation models
%that can accommodate large and negative 
%running \cite{Kawasaki:2003zv,Feng:2003mk}, only one model has been
%known to realize sizable negative running and a large tensor-to-scalar
%ratio simultaneously, making use of two or more trigonometric functions 
% \cite{Feng:2003mk} which is an extension of natural inflation 
%model \cite{Freese:1990rb}.

As usual, we incorporate tensor perturbations $h_{ij}$ to spatially flat 
Friedmann-Lemaitre-Robertson-Walker (FLRW) spacetime as
\beq
 ds^2=-dt^2+a^2(t)(\delta_{ij}+h_{ij})dx^idx^j=
a^2(\tau)[-d\tau^2+(\delta_{ij}+h_{ij})dx^idx^j],
\eeq
where $a(t)$ is the scale factor, $\tau$ is the conformal time, 
and indices $i,j$ run from 1 to 3. We impose transverse-traceless
condition on $h_{ij}$, $\partial_i h_{ij}=0=h^i_i$.

We perform Fourier expansion as
\beq
 h_{ij}(t,\vect{x})=\sum_{\lambda=+,\times}^{}\int\frac{d^3k}{(2\pi)^{3/2}}\epsilon_{ij}^{\lambda}
    (\vect{k})h_{\vecs{k}}^{\lambda}(t)e^{i\vecs{k}\cdot\vecs{x}},
\eeq
where the polarization tensor $\epsilon_{ij}^{\lambda}~(\lambda=+,\times)$ is normalized
as 
$\sum_{i,j}^{}\epsilon_{ij}^{\lambda}\epsilon_{ij}^{*\lambda^{\prime}}=
2\delta^{\lambda\lambda^{\prime}}$.

Since each polarization mode satisfies 
\beq
  \ddot{h}_{\vecs{k}}^{\lambda}(t)+3H\dot{h}_{\vecs{k}}^{\lambda}(t)+
 \frac{k^2}{a^2(t)}h_{\vecs{k}}^{\lambda}(t)=0,  \label{KG}
\eeq
which is the same as the Klein-Gordon equation for a massless
minimally coupled scalar field, we can quantize it during inflaton
using the conventional wisdom of quantum field theory in curved
spacetime \cite{staro}.  Here $H$ is the Hubble parameter and a dot
denotes time derivative.
As a result, we find a nearly scale-invariant long wave fluctuation
as an initial condition \cite{staro} with the amplitude
\beq
  \Delta_h^2(k)=\langle h_{ij}h^{ij}(k)\rangle  =64\pi G
\left( {\frac{{H(t_k) }}{{2\pi }}}\right)^2 ,
\eeq
where $H(t_k)$ is the Hubble parameter during inflation when
$k$-mode left the horizon, and the prefactor is determined by the
canonical quantization based on
 the Einstein action.  This weak wavenumber dependence may be
incorporated by Taylor expansion with respect to $\ln(k/k_{\piv})$ as
\begin{equation}
\Delta_h^2(k)=\Delta_h^2(k_{\piv})
\exp\left[n_{T}(k_{\piv})\ln\frac{k}{k_{\piv}}
+\frac{1}{2!}\alpha_{T}(k_{\piv})\lmk\ln\frac{k}{k_{\piv}}\rmk^2
+\frac{1}{3!}\beta_{T}(k_{\piv})\lmk\ln\frac{k}{k_{\rm
piv}}\rmk^3+\cdots\right].
\label{deltah}
\end{equation}
Here $k_{\piv}$ 
is a pivot scale where  the tensor-to-scalar
ratio $r(k_{\piv})$ is formally defined by
\beq
  r\equiv \frac{\Delta_h^2(k_{\piv})}{\Delta_s^2(k_{\piv})},
 \label{r}
\eeq
with $\Delta_s^2(k_{\piv})$ being the square amplitude of curvature
perturbation at the pivot scale $k_\ast $.
Planck takes $k_{\piv}=0.002{\rm Mpc}^{-1}$ \cite{Ade:2013zuv}. 
In a single-field slow-roll inflation model with a potential $V[\phi]$,
coefficients in (\ref{deltah}) as well as $r$ are given by the slow-roll
parameters,
\beq
  \epsilon_V[\phi]\equiv \frac{\mg^2}{2}\lmk\frac{V'[\phi]}{V[\phi]}\rmk^2,~~~
 \eta_V[\phi]\equiv \mg^2\frac{V''[\phi]}{V[\phi]},~~~\xi_V[\phi]\equiv \mg^4
 \frac{V'[\phi]V'''[\phi]}{V[\phi]^2},
\eeq
as
\beq
 r=16\epsilon_V,~~~n_T=-2\epsilon_V,~~~
\alpha_T(k)=-4\epsilon_V(2\epsilon_V-\eta_V),~~~  
\beta_T(k)=-4\epsilon_V(16\epsilon_V^2+2\eta_V^2-14\epsilon_V\eta_V+\xi^2_V),
\eeq
respectively.
% In \cite{Kuroyanagi:2011iw}, the Taylor expansion has been
% calculated up to sixth order, which provides an accurate amplitude
% of the spectrum at the direct detection scale $\sim 0.1$Hz.  For
% specific models, however, we do not use the above expression
% (\ref{deltah}) but numerically solve for $H(t_k)$ using full
% equations of motion.  In order to connect the number of $e$-folds of
% inflation with the frequency of gravitational waves, we numerically
% solve field equations
In order to obtain the accurate amplitude of gravitational waves
at the direct detection scale $\sim 1$Hz, (\ref{deltah}) is insufficient
and we must continue Taylor expansion  up to sixth order
\cite{Kuroyanagi:2011iw}. Then it agrees with the full numerical
solution of the mode equation for the models we consider here 
\cite{Kuroyanagi:2008ye,Kuroyanagi:2014qaa}.  In this paper, we use
 the sixth-order Taylor expansion
 \cite{Kuroyanagi:2011iw} in order to save the computation
time.   

The field equations for $\phi$ are given
as
\beq
 \ddot\phi +3H\dot\phi +V'[\phi]=0,~~\lmk\frac{\dot a}{a}\rmk^2
=\frac{\rho_\phi}{3\mg^2},~~\rho_\phi=\frac{1}{2}\dot\phi^2+V[\phi],
\eeq
during inflation, and
\beq
  \ddot\phi +(3H+\Gamma)\dot\phi
  +V'[\phi]=0,~~\frac{d\rho_r}{dt}=-4H\rho_r+\Gamma\dot\phi^2 ,~~
\lmk\frac{\dot a}{a}\rmk^2
=\frac{1}{3\mg^2}(\rho_\phi+\rho_r),
\eeq
in the field oscillation regime where $\rho_r$ is the radiation energy
density and $\Gamma$ is the decay rate of the inflaton that determines
the reheating temperature.  In the mid and late field oscillation regime
when the mass term dominates the potential,
the scalar field equation is replaced by
\beq
 \frac{d\rho_\phi}{dt}=-(3H+\Gamma)\rho_\phi.
\eeq

During slow roll inflation, $\ddot\phi$ and kinetic energy term are
negligible and we can find a well known slow-roll analytic solution
for the massive chaotic inflation \cite{chaoinf} with a quadratic potential 
\beq
V[\phi]=m^2\phi^2/2.  \label{chaopote}
\eeq
We can express the number of $e$-folds
$N\equiv \ln(a_{\rm end}/a)$ as
\begin{equation}
N \cong \frac{1}{M_{\rm Pl}^2}\int^{\phi}_{\phi_{\rm end}}
\frac{V[\phi]}{V^{\prime}[\phi]}d\phi=\frac{\phi^2}{4M_{\rm
Pl}^2}-\frac{1}{2},
~~~\phi_{\rm end}=\frac{\mpl}{\sqrt{2}}.
\end{equation}
Then the slow-roll parameters and $r$ are given in terms of $N$,
\begin{eqnarray}
\epsilon_V=\eta_V=2\frac{M_{\rm Pl}^2}{\phi^2}=\frac{1}{2N+1},~~~
r=16\epsilon_{V}=\frac{16}{2N+1}. \label{rchaotic} 
\end{eqnarray}
All the higher derivative slow-roll parameters including 
$\xi_V$
are equal to zero. 

From the numerical solution of field equations, we can calculate the
number of $e$-folds of inflation after the pivot scale $k_{\piv}$
left the Hubble radius as
\beq
 N_{\ast}=54.4+\frac{1}{3}\ln\lmk\frac{T_R}{10^8\rm GeV}\rmk.
 \label{efoldchaotic}
\eeq

From (\ref{rchaotic}) and (\ref{efoldchaotic}), we can obtain the
reheating temperature modulo dilution factor once the tensor-to-scalar
ratio is measured as
\beq
 % \frac{T_R}{F}
  T_R=1.7\times 10^6\exp\lkk
  160\lmk\frac{r}{0.15}\rmk^{-1}\rkk {\rm ~GeV}.  \label{tr}
%e^{160\lmk\frac{r}{0.15}\rmk^{-1}}
\eeq
 
%For another model with smaller $r$.....

Later each mode evolves as
\beq
 h_{\vecs{k}}^{\lambda}(t)\propto
a(t)^{\frac{1-3p}{2p}}J_{\frac{3p-1}{2(1-p)}}\lmk\frac{p}{1-p}
\frac{k}{a(t)H(t)}\rmk=a(\tau)^{\frac{1-3p}{2p}}
J_{\frac{3p-1}{2(1-p)}}(k\tau), \label{7}
\eeq
in a power-law background
 $a(t)\propto t^p$ with $p<1$.  Here $J_n(x)$ is a Bessel function.
  Thus the amplitude of
$h_{\vecs{k}}^{\lambda}(t)$ remains constant in the super-horizon regime
and it starts to decrease inversely proportional to the scale factor
after horizon crossing, so that the energy density of gravitational
waves,
\beq
  \rho_{\rm GW}=\frac{1}{64\pi Ga^2}\langle (\partial_{\tau}h_{ij})^2
+(\nabla h_{ij})^2 \rangle,
\eeq
decreases in the same way as radiation for modes well inside the horizon.

We define density parameter of gravitational waves of each logarithmic
frequency interval as
\beq
 \Omega_{\rm GW}(f,t_0)\equiv\frac{1}{\rho_c}\frac{d\rho_{\rm GW}}{d\ln k}
     =\frac{1}{12}\left(\frac{k}{a_0 H_0}\right)^2\Delta_h^2(k)T_h^2(f)
 %= \frac{h_F^2(f)}{16\pi G \rho_{cr}}
%\lmk\frac{2\pi fa_0}{a_{in}(f)}\rmk^2
%=\frac{1}{18\pi^2} \frac{V[\phi(f)]}{M_{G}^4}
= \frac{(2\pi f)^2}{12H_0^2}\Delta_h^2(f)T_h^2(f),
%\lmk\frac{a_{in}(f)}{a_0}\rmk^2. 
\label{infnorm}
\eeq 
where $T_h(k)$ is the transfer function given by
\beq
  T_h^2(k)=\Omega_m^2\frac{g_\ast(T_{\rm in})}{g_{\ast 0}}
 \lmk\frac{g_{\ast s0}}{g_{\ast}(T_{\rm in})}\rmk^{4/3}
\frac{9{j_1^2(k\tau_0)}}{(k\tau_0)^2}
T^2_1(x_{\rm eq})T_2^2(x_{\rm R}).
\eeq
Here the subscripts 0 and in refer to the values at
 the present time $\tau_0$ and at the  
epoch when $k$-mode reentered the Hubble radius, respectively.
$g_\ast$ and $g_{\ast s}$ represent 
the effective numbers of degrees of freedom for energy and
entropy densities with their current values
given by $g_{\ast 0}=3.36$ and $g_{\ast s0}=3.90$, respectively.
We take $\tau_0=2H_0^{-1}$, which is the right expression
for matter domination, and incorporate the effect of dark energy
by the factor $\Omega_m=1-\Omega_{DE}$. 
In the limit $k\tau_{0} \gg 1$ in which  we are interested, 
we can replace the square of the spherical Bessel function by averaging
over the period as $\overline{j_1^2(k\tau_0)}\cong (2k\tau_0)^{-2}$.
The first transfer function $T_1(x_{\rm eq})$ represents the effect of
transition from the radiation to the matter domination
\cite{Turner:1993vb}, 
while the second
one  $T_2(x_{\rm R})$ stands for that from the field oscillation regime
to the radiation dominated era at the reheating just after inflation
\cite{Nakayama:2008wy}.
They are explicitly given by \cite{Jinno:2013xqa}
\beq
  T_1^2(x_{\rm eq})=(1+1.41x_{\rm eq}+3.56x_{\rm eq}^2),
     \label{trans1}
\eeq
with $x_{\rm eq}=k/k_{\rm eq}$ and $k_{\rm eq}\equiv\tau_{\rm
  eq}^{-1}=7.1\times 10^{-2}\Omega_m h^2\,{\rm Mpc}^{-1}$, and \cite{Kuroyanagi:2014nba}
\beq
 T_2^2(x_{\rm R})=(1-0.22x_{\rm R}^{1.5}+0.65x_{\rm R}^2)^{-1},
     \label{trans2}
\eeq
with  $x_{\rm R}=k/k_{\rm R}$ and $k_{\rm R}\simeq 1.7\times
10^{14}{\rm Mpc}^{-1}(g_{*s}(T_{\rm R})/106.75)^{1/6}(T_{\rm
  R}/10^7{\rm GeV})$, respectively. 
Here $k_{\rm R}$ is related with the current frequency
\beq
   f_{\rm R}=\frac{k_{\rm R}}{2\pi a_0}\simeq 0.26{\rm Hz}
     \left(\frac{g_{*s}(T_{\rm R})}{106.75}\right)^{1/6}
     \left(\frac{T_{\rm R}}{10^7{\rm GeV}}\right).   
     \label{f_R}
\eeq
This is the frequency where the spectrum of stochastic gravitational wave
background is bent and the reheating temperature is encoded.
The density parameter therefore
 behaves as $\Omega_{\rm GW}(f,t_0)\propto f^{-2} (f^0)$
for the mode which enters the horizon 
in the matter (radiation) dominated regime.
More generally, for modes entering the horizon when the
equation-of-state parameter or the ratio of pressure to energy density
was $w$, its frequency dependence reads
\beq
  \Omega_{\rm GW}(f)\propto f^{\frac{2(3w-1)}{3w+1}}.
\eeq

The above is the case where the Universe evolves adiabatically after reheating
at the end of inflation.  If there is additional entropy production 
from a decaying matter component $\chi$ like Polonyi \cite{Moroi:1994rs}
or moduli fields \cite{Moroi:1999zb}
which temporarily dominates
the cosmic energy density, short-wave gravitational radiation is diluted
and $k_{\rm R}$ is modified \cite{Seto:2003kc,Nakayama:2008wy}.
In terms of the dilution factor,
\beq
  	F =\frac{s(T_\chi)a^3(T_\chi)}{s(T_R)a^3(T_R)} ,
\eeq
we find the transfer function in this case as
\beq
 T_h^2(k)=\Omega_m^2\frac{g_\ast(T_{\rm in})}{g_{\ast 0}}
 \lmk\frac{g_{\ast s0}}{g_{\ast}(T_{\rm in})}\rmk^{4/3}
\frac{9{j_1^2(k\tau_0)}}{(k\tau_0)^2}
T^2_1(x_{\rm eq})T_2^2(x_{F \rm R})
T^2_3(x_{\chi})T_2^2(x_{\chi R}),
\eeq
where \cite{Kuroyanagi:2014nba}
\beq
 T_3^2(x_{\chi})=(1+0.59x_{\chi}+0.65x_{\chi}^2),
\eeq
and
$x_{F\rm R},~x_{\chi},$ and $x_{\chi R}$ are given by
\beq
 x_{F\rm R}=\frac{F^{1/3}k}{k_R},~~
 x_{\chi}=\frac{k}{k_\chi},~~
 x_{\chi R}= \frac{k}{F^{2/3}k_\chi},
\eeq
with
\beq
k_\chi= 1.7\times 10^{7}~{\rm Mpc^{-1}}
	\left ( \frac{g_{*s}(T_\chi)}{106.75} \right )^{1/6}
	\left ( \frac{T_\chi}{1~{\rm GeV}} \right ).  \label{k_chi}
\eeq
The Universe is dominated by $\chi$ in the regime modes with
$k_\chi < k < F^{2/3}k_\chi$ enter the horizon.
For wavenumber $k_{\chi R}<k<k_R$, which corresponds 
to the mode entering the horizon in the radiation dominated era 
before $\chi$-domination, the energy density of the gravitational waves 
is suppressed by the factor 
$\sim (k_\chi/k_{\chi R})^2= F^{-4/3}$ \cite{Seto:2003kc}.
We also note that the frequency where the spectrum bends due to the reheating
is modified as
\beq
  	f_{F \rm R}= 0.026~{\rm Hz}
	F^{-1/3}\left ( \frac{g_{*s}(T_R)}{106.75} \right )^{1/6}
	\left ( \frac{T_R}{10^6~{\rm GeV}} \right ),  \label{f_RF}
\eeq
as is clear from the above discussion.
Furthermore in the presence of late-time entropy production
(\ref{efoldchaotic}) and (\ref{tr}) are modified as
\beq
 N_{\ast}=54.4+\frac{1}{3}\ln\lmk\frac{T_R/F}{10^8\rm GeV}\rmk,
 \label{efoldchaoticF}
~~~
 \frac{T_R}{F}
  =1.7\times 10^6\exp\lkk
  160\lmk\frac{r}{0.15}\rmk^{-1}\rkk {\rm ~GeV},  
%e^{160\lmk\frac{r}{0.15}\rmk^{-1}}
\eeq
respectively.

We analyze the detectability of $T_R$  in terms of
the Fisher information matrix approach following
\cite{Kuroyanagi:2011fy} (see also
\cite{Seto:2005qy,Fisherref,Jinno:2014qka}),
and discuss improvements of specifications of DECIGO to increase the
measurable range of $T_R$.

Suppose that the stochastic gravitational wave background 
$\Omega_{\rm GW}(f,t_0)$ depends on parameters $p_i$.  Then since its
detection is done based on cross correlation of two (or more) detectors,
the Fisher matrix element ${\cal F}_{ij}$ is determined by their noise
power spectra $N_I(f)$ as
\begin{equation}
     {\cal F}_{ij}=\left(\frac{3H_0^2}{10\pi^2}\right)^2 2\,T_{\rm obs}\nonumber\\
     \sum_{(I,J)}\int^{f_{\rm max}}_{f_{\rm cut}}df\frac{|\gamma_{IJ}(f)|^2\partial_{p_i}
     \Omega_{\rm GW}(f)\partial_{p_j}\Omega_{\rm GW}(f)}{f^6N_I(f)N_J(f)},
     \label{Fisher}
\end{equation}
where $T_{\rm obs}$ is observation time and 
$\gamma_{IJ}(f)$ is the overlap reduction function determined by
the survey configuration \cite{Allen:1997ad}.  
Here we take $f_{\rm cut}=0.1$Hz unless otherwise
stated, below 
which the signal may be contaminated by
noise from cosmological white dwarf binaries \cite{Farmer:2003pa}.
We set $f_{\rm max}=\infty$, 
but this choice is largely irrelevant to the final result since
 the high-frequency range is limited by the noise
spectrum as shown in \cite{Kuroyanagi:2011fy}.

We consider two sets of Fabry-Perot type DECIGO (FP-DECIGO) each
consisting of three satellites with an equilateral triangular
configuration.  The overlap reduction function is given as
\beq
  \gamma_{IJ}(f)=\frac{5}{8\pi}\int d\vect\Omega e^{i2\pi f
    \vecs\Omega\cdot
    \Delta\vect{x}}\sum_{A=+,\times}F_I^A(f,\vect\Omega)F_J^A(f,\vect\Omega),
\eeq 
where $\Delta\vect{x}$ is the separation between the two detectors and
$F_I^A(f,\vect\Omega)$ is the detector pattern functions. For details,
see e.g. \cite{Maggiore:1999vm}.

The noise power spectrum consists of three mutually independent major
components, namely, laser shot noise, $S_{\rm shot}$, radiation pressure
noise, $S_{\rm rad}$, and acceleration noise, $S_{\rm acc}$.
The noise spectrum of two sets of identical detectors is given by
\begin{equation}
N_1(f)=N_2(f)=S_{\rm shot}(f)+S_{\rm rad}(f)+S_{\rm acc}(f).
\end{equation}
As with the ground-based Fabry-Perot type interferometers, they depend
on the arm length $L$, the laser power, $P$, the laser wavelength,
$\lambda$, the mirror mass, $M$, the finesse, $\calF$ etc.  
The shot noise and the radiation pressure noise are given by
\begin{align}
S_{\rm shot}^{1/2}(f)&=\frac{\sqrt{\hbar\pi c\lambda}}{4\calF
 L\sqrt{\tilde{P}}}
\lkk 1+\lmk\frac{f}{f_c}\rmk^2\rkk^{1/2}, \\
S_{\rm rad}^{1/2}(f)&= \frac{16\calF}{(2\pi f)^2ML}\sqrt{\frac{\hbar
 P}{\pi\lambda c}}\lkk 1+\lmk\frac{f}{f_c}\rmk^2\rkk^{-1/2},
\end{align}
respectively, where $c$ and $\hbar$ are explicitly shown for clarity
\cite{maggiore}. 
Here the cut off frequency $f_c$ is given by
\beq
 f_c=\frac{c}{4L\calF}.
\eeq
The finesse, which is the ratio of the resonance separation to the
resonance width, and the effective laser output power $\tilde{P}$
are given by the reflectivities of the two mirrors, $r_F$ and $r_E$,
the former referring to the front mirror closer to the laser 
and the latter the end mirror,
and the transmutation rate $t_F$ of the front mirror 
as
\beq
  \calF=\frac{\pi\sqrt{r_Fr_E}}{1-r_Fr_E},~~~
\tilde{P}=\lmk\frac{t_F^2r_E}{1-r_Fr_E}\rmk^2P.
\eeq 
These parameters are related to $L,~\lambda$, and the mirror radius $R$
as follows.
\beq
 r_F=r_{Fm}r_{G},~~r_E=r_{Em}r_G,~~ t_F=\sqrt{r_G^2-r_{Fm}^2},~~
r_G=1-\exp\lmk -\frac{2\pi R^2}{\lambda L}\rmk.
\eeq
Here $r_G$ is the fraction of photons of the ideal Gaussian beam
 received by the mirror, and $r_{Fm}$ and $r_{Em}$ denote intrinsic
 reflectivity of each mirror.  We take $r_{Fm}^2=0.67$
and $r_{Em}^2=0.9999$ below.

The acceleration noise, on the other hand, is set to take smaller values
than the radiation pressure noise with the effective strain
\beq
  S_{\rm acc}^{1/2}(f)=\frac{16\calF}{2(2\pi f)^2LM}\sqrt{\frac{\hbar
  P}{\pi\lambda c}},
\eeq
that is, $S_{\rm rad}^{1/2}(f)/3$ at lower frequencies \cite{Kawamura:2011zz}.
Since the maximum measurable reheating temperature is determined by
sensitivities at higher frequency, the acceleration noise is actually
irrelevant in the subsequent discussion, and the dominant source of
detector noise in the relevant frequency range is the shot noise.

The original DECIGO assumes $L=10^3$km, $P=10$W, $\lambda=532$nm,
$M=100$kg, and $R=0.5$m.  With 3 year observations, it will achieve
the best sensitivity $\Omega_{\rm GW}=7.2\times 10^{-16}$ at
$f=0.2$Hz, corresponding to the effective strain $h=2.8\times
10^{-25}$.  Since this is not sufficient to remove foreground
contamination from neutron star binaries \cite{Yagi:2011wg}, upgrade
of specifications to achieve three times better sensitivity has been
discussed %\cite{3bai} 
which takes $L=1.5\times 10^3$km, $P=30$W,
$\lambda=532$nm, $M=100$kg, and $R=0.75$m.  Then the best sensitivity
$\Omega_{\rm GW}=3.2\times 10^{-16}$ is achieved at $f=0.5$Hz
corresponding to $h=3.0\times 10^{-26}$.

In order to keep the same level of sensitivity to $\Omega_{\rm GW}(f)$ at
higher frequencies, we must suppress the noise power spectrum
$N_I(f)$ in proportion to $f^{-3}$.  
To reduce the sensitivity at high frequencies,
the shot noise is more important than
the radiation pressure noise.  To suppress the former, we should
decrease the laser wavelength $\lambda$ or increase the power $P$,
$\calF$, and $L$.  However, as we increase $\calF L$, $f_c$ becomes
smaller and the frequency range of our interest falls above $f_c$
where $S_{\rm shot}(f)$ does not depend on $\calF$ nor $L$.
Thus the magnitude of the shot noise can only be controlled by
$\lambda$ and $\tilde P$ in the relevant range.  

\begin{figure}
 \begin{center}
  \includegraphics[width=0.45\textwidth]{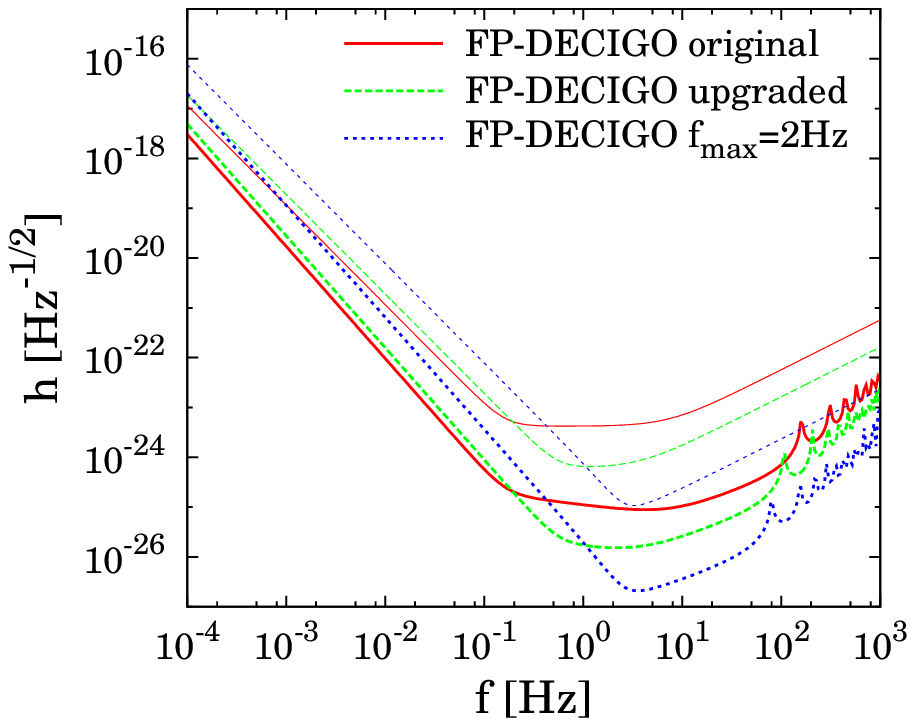}
  \includegraphics[width=0.45\textwidth]{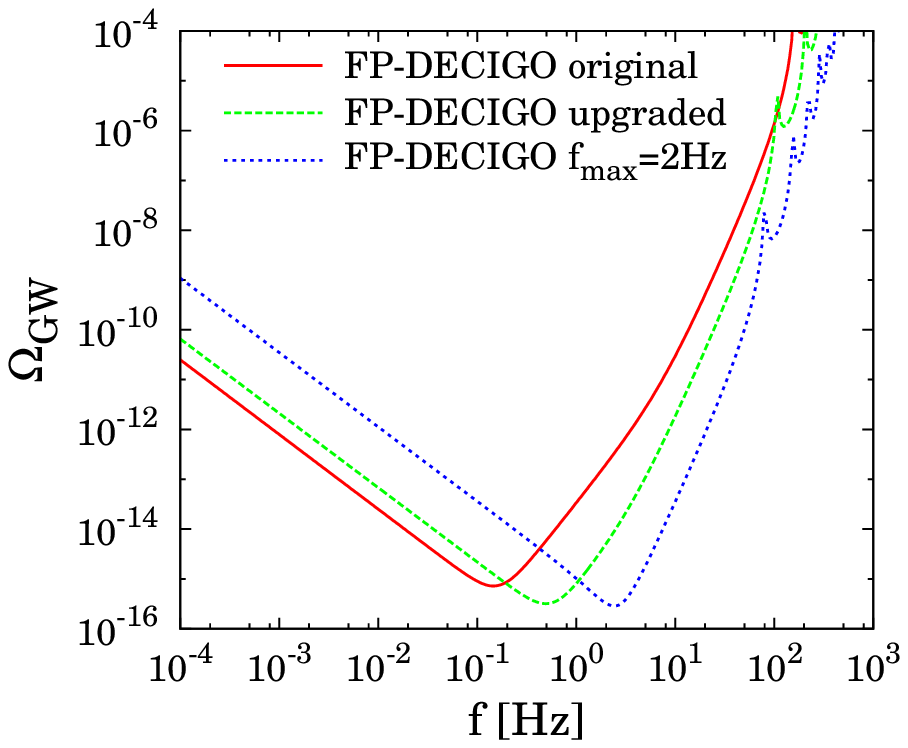}
   \caption{\label{Fig:noise_curve} 
Noise curves of three types of FP-DECIGO shown in terms of 
the strain (left panel) and the
  detectability of $\Omega_{\rm GW}$ for 3 years of observations (right panel).
In the left panel, we show the strain sensitivity for one set of detectors (thin curves) and the effective strain sensitivity obtained assuming a 3-year cross-correlation analysis between two sets of detectors (thick curves).  In the right panel, we only show the sensitivity curves for the cross-correlation analysis.
}
 \end{center}
\end{figure}

Of course, $\lambda$ cannot be set arbitrary small.  Let us tentatively
take a deep UV wavelength $\lambda=157$nm. Lasers with this wavelength
and mirrors with high reflectivity for this band
are commercially available now, 
but stability and high enough power are challenges to be
solved by the time when DECIGO will be put into practice in the coming decades.
With $\lambda=157$nm and the laser power $P=300$W, we could achieve a sensitivity of
$\Omega_{\rm GW}$ at $f=2$Hz similar to that of the upgraded DECIGO at
$f=0.5$Hz.  
If we should stick to $\lambda=532$nm, we could obtain the same noise
curve by boosting the power to $P\simeq 1$kW since 
both $S_{\rm shot}(f)$ and $S_{\rm rad}(f)$ depend on the combination
$P/\lambda$ only.  In this case, we must take $R=1.38$m to preserve $r_G$.
Hereafter we shall call this specification the {\it upshifted} DECIGO and
label it  as $f_{\max}=2$Hz in the figures to avoid confusion.
Figures \ref{Fig:noise_curve} represent sensitivity of three 
types of FP-DECIGO 
in terms of the strain (left panel) and $\Omega_{\rm GW}(f)$ of
the stochastic gravitational wave background for 3 years of
observation.

%%%%%%%%%%%%%%%%%%%%
%%%%%%%%%%%%%%%%%%%%%

\begin{figure}[h]
 \begin{center}
 \includegraphics[width=0.45\textwidth]{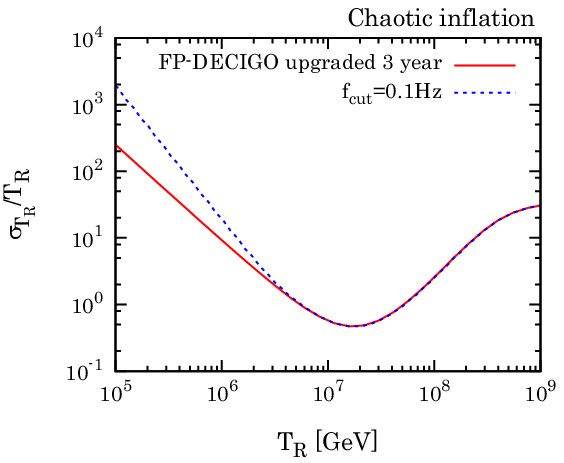} 
\includegraphics[width=0.45\textwidth]{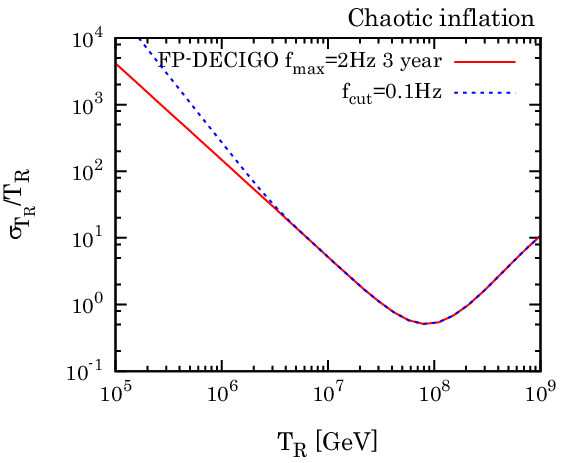}
  \caption{\label{Fig:TRchaotic} The marginalized one $\sigma$
    uncertainty in $T_{\rm R}$ as a function of $T_{\rm R}$ for
    the upgraded FP-DECIGO (left panel) and the 
upshifted FP-DECIGO with the maximal
    sensitivity set to $f_{\max}=2$Hz (right panel).  Red
    solid line represents the case with no foreground contamination
and blue dotted line shows the case where the range
  $f<f_{\rm cut}=0.1$Hz is fully contaminated by binary white dwarfs
and cannot be used in the analysis.}
 \end{center}
\end{figure}

Using the above set up, let us first consider the case of chaotic
inflation \cite{chaoinf} with a quadratic potential (\ref{chaopote}).
From Figs.\ \ref{Fig:TRchaotic}, we find that the maximum value of the
reheating temperature that can be determined with $\sigma_{T_R}<T_R$
is $T_R=5.2\times 10^{7}$GeV for the upgraded DECIGO and
$T_R=2.1\times 10^8$GeV for the upshifted DECIGO $f_{\max}=2$Hz, where
$\sigma_{T_R}^2$ is the variance of the measurement error.  Thus by
improving the specification, we can improve the upper limit of the
measurable reheating temperature by a factor of 4.  We also note that
the low frequency cutoff at $f_{\rm cut}=0.1$Hz due to contamination
of white dwarf binaries does not affect the sensitivity to measure
$T_R$ as long as we pay attention to the range $\sigma_{T_R}< T_R$.
Here we have taken Planck+WP+highL+BAO mean values of cosmological
parameters, $\Omega_{\Lambda}=0.692$, $\Omega_m h^2=0.141$, $h=0.678$
and $\Delta_s^2=2.21\times 10^{-9}$ at $k=0.05{\rm Mpc}^{-1}$ which
corresponds to $\Delta_s^2=2.51\times 10^{-9}$ at our pivot scale
$k_{\piv}=0.002{\rm Mpc}^{-1}$ for $n_s=0.9608$ \cite{Ade:2013zuv}.

%%%%%%%%%%%%%%%%%%%%
%%%%%%%%%%%%%%%%%%%%%

Since the combination of $\lambda$ and $P$ is already ambitious
enough, we cannot hope to go further to achieve the desired
sensitivity of $\Omega_{\rm GW}$ at even higher frequencies.  Because
we need the strain sensitivity $S_h$ must be improved in proportion to
$f^{-3}$, it is really difficult to accurately probe almost
scale-invariant stochastic gravitational wave background at higher
frequencies.  Just for completeness, we also consider the ideal case
of the ultimate-DECIGO whose noise curve is determined by the quantum
limit, with the following specifications: $L=5\times 10^5$km,
$M=$100kg, $\lambda=532$nm, $P=10$MW, and $R=3$m, and compare with the
above results.

%%%%%%%%%%%%%%%%%%%%
\begin{figure}[h]
 \begin{center}
  \includegraphics[width=0.45\textwidth]{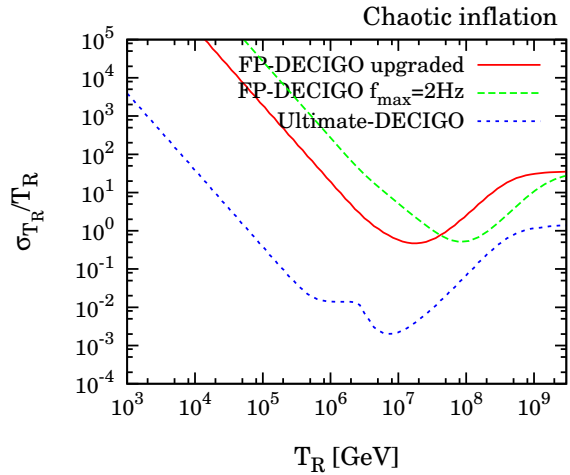}
  \caption{\label{Fig:TRchaoticult}   The marginalized
      one $\sigma$ uncertainty in $T_{\rm R}$ as a function of $T_{\rm
        R}$ for the Ultimate-DECIGO in comparison with the upgraded
      FP-DECIGO and the upgraded FP-DECIGO with the maximal sensitivity set to
      $f_{\max}=2$Hz.  The low frequency cutoff at $f_{\rm
        cut}=0.1$Hz due to contamination of white dwarf
      binaries is taken into account.  }
 \end{center}
\end{figure}
%%%%%%%%%%%%%%%%%%%%%

Figure\ \ref{Fig:TRchaoticult} depicts the uncertainty in the
measurement of $T_R$ in three specifications of DECIGO.
For this particular inflation model one can determine the reheating
temperature  by using the relation between the amplitude of
tensor perturbation and $N$ through (\ref{rchaotic}) and 
(\ref{efoldchaotic}) 
even without observing the spectral bend
due to the change of the equation of state during reheating regime.
This is why $\sigma_{T_R}$ saturates at higher $T_R$ for which
$f_R$ is beyond the observable range.  In practice,  we
do not know the exact shape of the inflaton potential a priori, hence
this saturation should be regarded as an artifact of our analysis using
a specific model of inflation from the beginning.
Fortunately, however, our results of the measurable ranges of $T_R$
within one $\sigma_{T_R}$ accuracy are not affected by this as seen 
in Fig.\ \ref{Fig:TRchaoticult}, whose numerical values are shown in 
Table I.

\begin{table}[h]
\begin{tabular}{lcc}
Chaotic inflation & lower limit & upper limit \\
\hline 
FP-DECIGO upgraded &  ~~~$5.6\times 10^6$ GeV~~~ &  $5.2\times 10^7$ GeV \\
FP-DECIGO $f_{\max}$=2Hz &   ~~~$3.3\times 10^7$ GeV~~~ &  $2.1\times 10^8$ GeV \\
Ultimate-DECIGO&    ~~~$6.2\times 10^4$ GeV~~~ &  $7.0\times 10^8$ GeV    \\
\hline 
\end{tabular}
\caption{Range of the reheating temperature $T_R$ which can be measured
 with each specification of DECIGO for massive scalar chaotic inflation.}
\end{table}

\begin{figure}
 \begin{center}
 \includegraphics[width=0.45\textwidth]{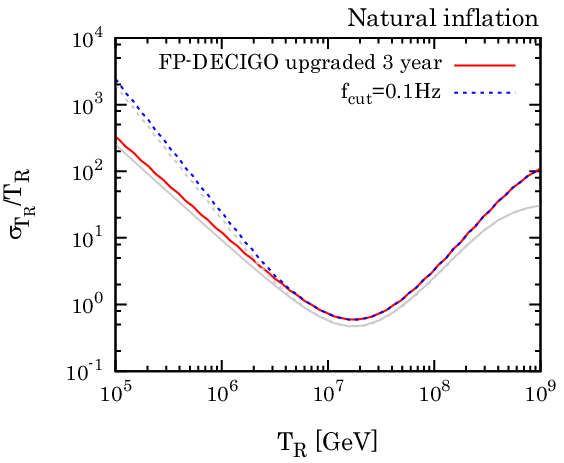}
  \includegraphics[width=0.45\textwidth]{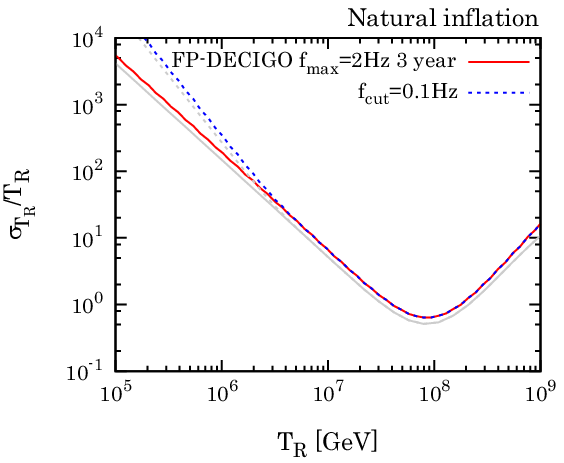}
  \caption{\label{Fig:TRnatural} The marginalized $1\sigma$
    uncertainty in $T_{\rm R}$ as a function of $T_{\rm R}$ for
    natural inflation with $v=7\mpl$.  The left panel is for the
    upgraded FP-DECIGO and the right panel is for the upshifted
    FP-DECIGO with the maximal sensitivity set to $f=2$Hz.  For
    comparison, the results for chaotic inflation are also plotted by
    gray curves.  }
 \end{center}
\end{figure}

\begin{figure}
 \begin{center}
  \includegraphics[width=0.45\textwidth]{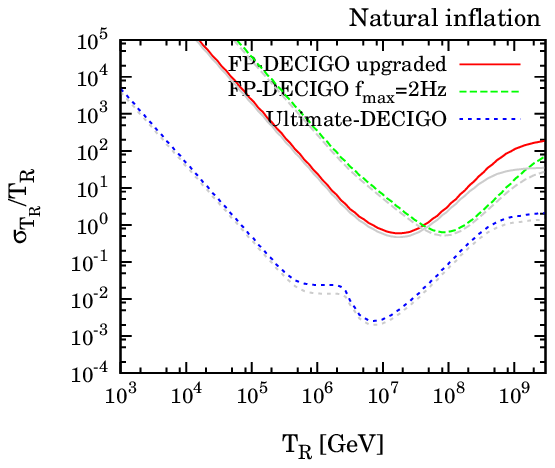}
  \caption{\label{Fig:TRnaturalult} The comparison of marginalized
    $1\sigma$ uncertainty in $T_{\rm R}$ for natural inflation with
    $v=7\mpl$.  The low frequency cutoff at $f_{\rm cut}=0.1$Hz due to
    contamination of white dwarf binaries is taken into account. }
 \end{center}
\end{figure}

Next we consider natural inflation \cite{Freese:1990rb} with a potential
\begin{equation}
V[\phi]=\Lambda^4\left[1-\cos\left(\frac{\phi}{v}\right)\right].
\end{equation}
As a specific example, we take $v=7\mpl$ for which (\ref{efoldchaotic})
practically applies \cite{Kuroyanagi:2014qaa}.
In this model, the tensor-to-scalar ratio $r$ corresponding to the number
of $e$-folds $N$ is given by
\beq
  r=\frac{16\nu}{(1+\nu)e^{2\nu N}-1},~~~\nu\equiv \frac{\mpl^2}{2v^2},
\eeq
so we find $r=0.091$ for $N=50$ and $r=0.067$ for $N=60$.
In Figs. \ref{Fig:TRnatural} and \ref{Fig:TRnaturalult}, detectability
of $T_R$ in this model is depicted.  Saturation at higher $T_R$ should
again be regarded as an artifact, but it does not affect the one
$\sigma_{T_R}$ range, which is shown in Table II.  
As is seen there, as far as the upper
limit is concerned, the improvement from upgraded FP-DECIGO to
the upshifted FP-DECIGO
is larger than that from  the upshifted specification to
ultimate DECIGO, although the lower limit also shifts to a higher value
in the former case.
     
\begin{table}[h]
\begin{tabular}{lcc}
Natural inflation ($v=7\mpl$) & lower limit & upper limit \\
\hline 
FP-DECIGO upgraded &     ~~~$7.0\times 10^6$ GeV ~~~&  $4.2\times 10^7$ GeV \\
FP-DECIGO $f_{\max}$=2Hz &  ~~~$4.0\times 10^7$ GeV ~~~& $1.8\times 10^8$ GeV \\
Ultimate-DECIGO &       ~~~$6.9\times 10^4$ GeV ~~~&  $4.6\times 10^8$ GeV \\
\hline \\
\end{tabular}
\caption{Range of the reheating temperature $T_R$ which can be measured
 with each specification of DECIGO for natural inflation model with $v=7\mpl$.}
\end{table}

%%%%%%%%%%%%%%%%%%%%
%%%%%%%%%%%%%%%%%%%%%

%%%%%%%%%%%%%%%%%%%%
%%%%%%%%%%%%%%%%%%%%%

\begin{figure}
 \begin{center}
  \includegraphics[width=0.32\textwidth]{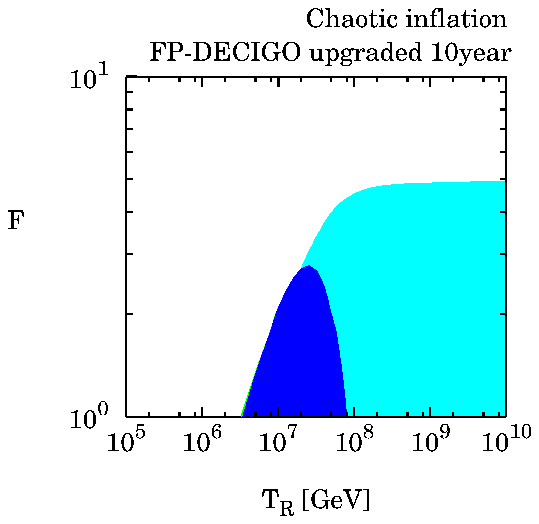}
  \includegraphics[width=0.32\textwidth]{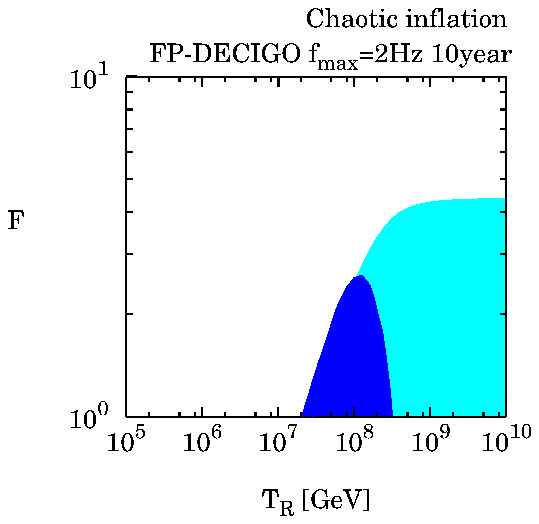}
  \includegraphics[width=0.32\textwidth]{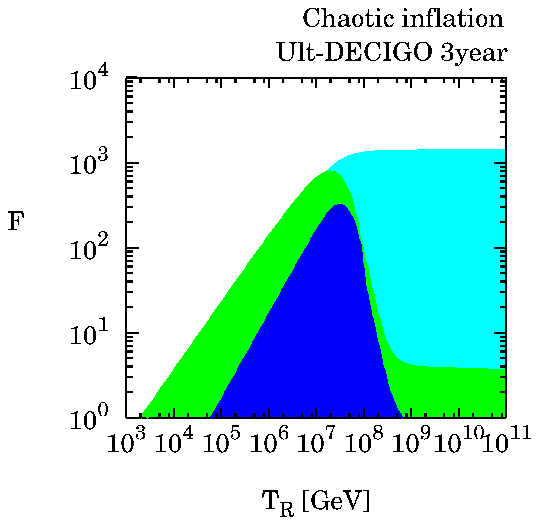}
  \caption{\label{Fchao} Light blue region shows the parameter range
    where stochastic gravitational wave can be detected with S/N$>1$
    for chaotic inflation of massive scalar potential.  Thick blue
    region represents the region the reheating temperature can be
    measured with an error less than  one $\sigma$ with the cutoff at
      $0.1$Hz, and green region describes the case without the cutoff.
      For both cases, information from the CMB is combined.  Note
      that, for FP-DECIGO, the thick blue region completely overlaps
      with the green region.  }
 \end{center}
\end{figure}

\begin{figure}
 \begin{center}
  \includegraphics[width=0.32\textwidth]{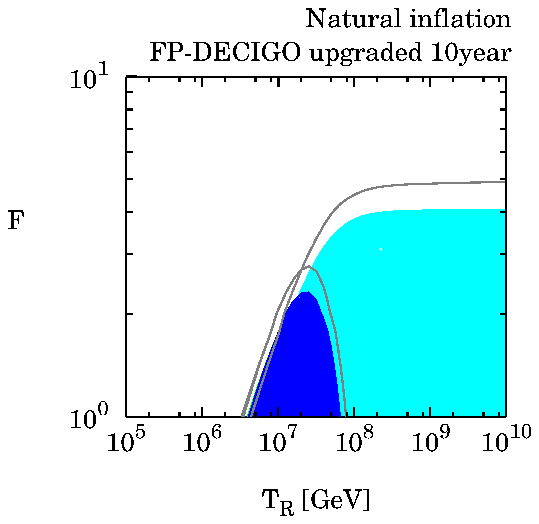}
  \includegraphics[width=0.32\textwidth]{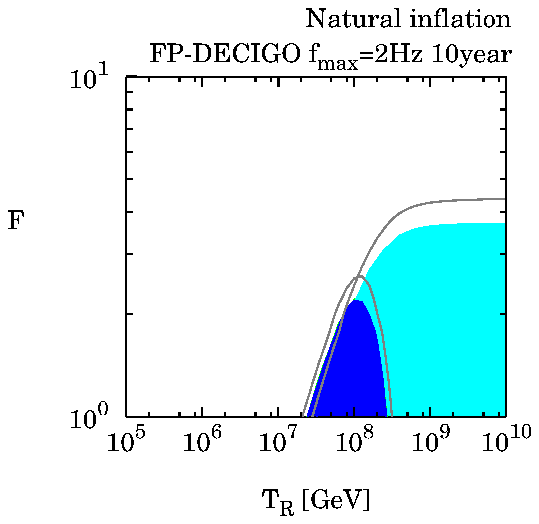}
  \includegraphics[width=0.32\textwidth]{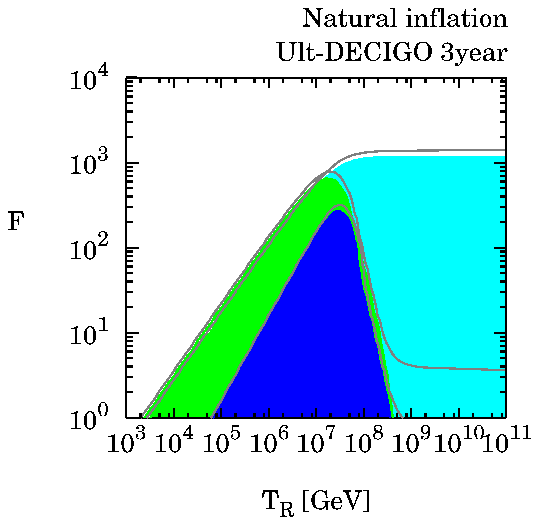}
  \caption{\label{Fnat} The same figure for natural inflation. For comparison, the results for chaotic inflation is also plotted by gray curves. }
 \end{center}
\end{figure}

Next we consider the case additional entropy production occurs after
reheating stage.  Figures \ref{Fchao} depict detectability of
gravitational waves in the case $F$ is larger than unity in chaotic
inflation.  Figures \ref{Fnat} are for the case of the natural
inflation with $v=7\mpl$.  As is seen there if $F$ is significant,
only ultimate DECIGO can measure the inflationary tensor perturbation.

Let us summarize the results so far.  The upgraded DECIGO and the
upshifted DECIGO with $f_{\max}=2$Hz can measure the reheating
temperature if it takes a value within a factor $\lesssim 10$ from
$5\times 10^6$GeV or $3\times 10^7$GeV, respectively.  With an
appreciable amount of entropy production it would be formidable to
detect the tensor perturbation itself, not to mention the
detectability of $T_R$ without ultimate DECIGO.  Hence let us assume
$F=1$ below and consider the implication of the range of $T_R$
measurable by the upgraded or upshifted DECIGO.

First of all, a large enough $r$ to make direct detection of
inflationary gravitational waves by DECIGO possible means high 
scale  inflation.  Thus one may naively think that the reheating
temperature is also high, which may not be the case as we argue now.  
As discussed in \cite{Lyth}, high scale
inflation means large excursion of the inflaton well beyond $\mpl$.
To stabilize the inflaton potential over such a large field range
we need some kind of symmetry, which may also constrain the coupling
between the inflaton and matter fields to delay reheating.  

The simplest example is natural inflation which is based on the
Nambu-Goldstone symmetry.  It is natural to suppose that matter
coupling with the inflaton is also protected by this symmetry and
typical coupling of the inflaton is suppressed by $v^{-1}$, so that on
dimensional grounds the decay rate of the inflaton reads
\beq
  \Gamma \approx g^2\frac{M^3}{v^2}\approx g^2\frac{\Lambda^6}{v^5},
\eeq
where $M\equiv \Lambda^2/v$ is the inflaton mass at 
the origin \cite{Freese:1990rb}.
This yields $T_R$ as
\beq
  T_R \approx 5\times 10^7\lmk\frac{g}{0.1}\rmk {\rm GeV}
\eeq
for $v=7\mpl$ where $g$ is a coupling constant. 

Chaotic inflation, on the other hand, is also naturally realized
introducing a shift symmetry $\phi\rightarrow \phi+iC$ with $C$
being a real number \cite{Kawasaki:2000yn}.
For example, let us take the K\"ahler potential and superpotential as~
\begin{gather}
	K  = \frac{1}{2}(\phi+\phi^\dagger)^2 +|X|^2 + |H_u|^2 + |H_d|^2, \\
	W = mX\phi + y X H_u H_d,
\end{gather}
where $\phi$ denotes the inflaton superfield whose imaginary component is regarded as the inflaton,
$X$ is a singlet chiral superfield and $H_u$ and $H_d$ are up- and down-type Higgs doublets, respectively.
In this minimal setup, the reheating takes place through the inflaton decay into the Higgs bosons and higgsinos and the
reheating temperature is given by
\begin{equation}
	T_{\rm R} \simeq 4\times 10^8\,{\rm GeV}\left( \frac{|y|}{10^{-6}} \right)\left( \frac{m}{10^{13}\,{\rm GeV}}\right)^{1/2}.
\end{equation}
Importantly, the smallness of the coupling constant $|y| \lesssim 10^{-6}$ is required from the successful inflaton dynamics
so that the Higgs fields should not become tachyonic during inflation~\cite{Nakayama:2013txa}.

%%%%%%%%%%%%%%%%%%%%
%%%%%%%%%%%%%%%%%

%%%%%%%%%%%%%%%%%%%%
%%%%%%%%%%%%%%%%%

Finally, we comment on the particle-physics aspects
 to determine the reheating temperature.
In supersymmetric (SUSY) theories, the abundance of the gravitino 
is proportional to the reheating temperature~\cite{Bolz:2000fu}:
\begin{equation}
	\frac{n_{3/2}}{s} \sim 2\times 10^{-12}\left( 1+\frac{m_{\tilde g}^2}{3m_{3/2}^2} \right)\left( \frac{T_{\rm R}}{10^{10}\,{\rm GeV}}\right),
\end{equation}
where $n_{3/2}$ and $s$ are the gravitino number density and the entropy density, $m_{3/2}$ $(m_{\tilde g})$ denotes the
gravitino (gluino) mass.
Hence the reheating temperature is bounded from above depending on the gravitino mass~\cite{Moroi:1993mb,Kawasaki:2004qu}.
It is interesting that the mass of the Higgs boson discovered at the LHC may imply relatively high scale SUSY~\cite{Ibe:2011aa,ArkaniHamed:2012gw,Arvanitaki:2012ps} of $m_{3/2} \sim 100-1000$\,TeV.
In this scenario, the upper bound on $T_{\rm R}$ reads $T_{\rm R}\lesssim 10^9-10^{10}$\,GeV so that the 
Winos produced by the gravitino decay should not be overabundant.\footnote{
	In this mass range of the gravitino, nonthermal gravitino production from the inflaton decay is also harmless~\cite{Nakayama:2014xca}.
}
This upper bound is close the sensitive range of the DECIGO as shown above.
If the upgraded LHC discovers supersymmetry in near
future, we will be able to obtain more useful information on the
reheating temperature from these particle physics considerations.
Needless to say, further observations of CMB temperature anisotropy and
B-mode polarization in near future will also provide us with invaluable
information on the tensor amplitude at the DECIGO band as well as to
determine the model of inflation.

Hence we should watch these progress to design the specifications of
DECIGO to extract maximal cosmological information, so that it can
determine when the Big Bang happened.

\acknowledgements 
We would like to thank Masaki Ando for providing a code to calculate the
sensitivity curves, Masatake Ohashi and Atsushi Taruya for useful
communications on the sensitivities of DECIGO.
This work was supported in part by JSPS Grant-in-Aid for Scientific Research
 (A) (No.26247042 [KN]), Scientific Research (B) (No.\ 23340058 [JY]), 
Young Scientists (B) (No.26800121 [KN])
and Grant-in-Aid for Scientific Research on Innovative Areas 2603 ([KN]).


\begin{thebibliography}{99}
\bibitem{oriinf2}
K.\ Sato, 
Mon.\ Not.\ Roy.\ Astron.\ Soc.  {\bf 195}, 467 (1981);
A.H.\ Guth, 
{ Phys.\ Rev.} {\bf D23}, 347 (1981)

\bibitem{oriinf1}
A.A.\ Starobinsky
Phys.\ Lett. {\bf 91B}, 99 (1980)

\bibitem{yuragi}
V.~F.~Mukhanov and G.~V.~Chibisov,
  %``Quantum Fluctuation and Nonsingular Universe. (In Russian),''
  JETP Lett.\  {\bf 33}, 532 (1981)
  [Pisma Zh.\ Eksp.\ Teor.\ Fiz.\  {\bf 33}, 549 (1981)];
S.W.\ Hawking, 
{ Phys.\ Lett.} {\bf 115B}, 295 (1982);
A.A.\ Starobinsky, 
Phys.\ Lett. {\bf 117B}, 175 (1982);~
A.H. Guth and S-Y. Pi, 
{ Phys.\ Rev.\ Lett.} {\bf 49}, 1110 (1982).
%

\bibitem{WMAP}
D.~N.~Spergel {\it et al.}  [WMAP Collaboration],
  %``First year Wilkinson Microwave Anisotropy Probe (WMAP) observations: Determination of cosmological parameters,''
  Astrophys.\ J.\ Suppl.\  {\bf 148}, 175 (2003)
  [astro-ph/0302209];
H.~V.~Peiris {\it et al.}  [WMAP Collaboration],
  %``First year Wilkinson Microwave Anisotropy Probe (WMAP) observations: Implications for inflation,''
  Astrophys.\ J.\ Suppl.\  {\bf 148}, 213 (2003)
  [astro-ph/0302225].
\bibitem{WMAP9}
G.~Hinshaw {\it et al.}  [WMAP Collaboration],
  %``Nine-Year Wilkinson Microwave Anisotropy Probe (WMAP) Observations: Cosmological Parameter Results,''
  Astrophys.\ J.\ Suppl.\  {\bf 208}, 19 (2013)
  [arXiv:1212.5226 [astro-ph.CO]].
\bibitem{Ade:2013zuv} 
  P.~A.~R.~Ade {\it et al.}  [Planck Collaboration],
  %``Planck 2013 results. XVI. Cosmological parameters,''
  arXiv:1303.5076 [astro-ph.CO].
\bibitem{Ade:2013uln} 
  P.~A.~R.~Ade {\it et al.}  [Planck Collaboration],
  %``Planck 2013 results. XXII. Constraints on inflation,''
  arXiv:1303.5082 [astro-ph.CO].
\bibitem{Ade:2013ydc} 
  P.~A.~R.~Ade {\it et al.}  [Planck Collaboration],
  %``Planck 2013 Results. XXIV. Constraints on primordial non-Gaussianity,''
  arXiv:1303.5084 [astro-ph.CO].
\bibitem{bicep}
P.~A.~R.~Ade {\it et al.}  [BICEP2 Collaboration],
 %``Detection of B-Mode Polarization at Degree Angular Scales by BICEP2,''
  Phys.\ Rev.\ Lett.\  {\bf 112}, 241101 (2014)
  [arXiv:1403.3985 [astro-ph.CO]].
\bibitem{staro}A.A.\ Starobinsky, JETP Lett. {\bf 30}, 682 (1979);
V.A.\ Rubakov, M.V.\ Sazin, and A.V.\ Veryaskin, Phys.\
 Lett. {\bf 115B}, 189(1982);
L.F.\ Abbott and M.B.\ Wise, Nucl.\ Phys.\
 B {\bf 244}, 541 (1984).

\bibitem{CervantesCota:1995tz} 
  J.~L.~Cervantes-Cota and H.~Dehnen,
  %``Induced gravity inflation in the standard model of particle physics,''
  Nucl.\ Phys.\ B {\bf 442}, 391 (1995)
  [astro-ph/9505069];
%\bibitem{Bezrukov:2007ep} 
  F.~L.~Bezrukov and M.~Shaposhnikov,
  %``The Standard Model Higgs boson as the inflaton,''
  Phys.\ Lett.\ B {\bf 659}, 703 (2008)
  [arXiv:0710.3755 [hep-th]];
%\bibitem{Barvinsky:2008ia} 
  A.~O.~Barvinsky, A.~Y.~.Kamenshchik and A.~A.~Starobinsky,
  %``Inflation scenario via the Standard Model Higgs boson and LHC,''
  JCAP {\bf 0811}, 021 (2008)
  [arXiv:0809.2104 [hep-ph]].
\bibitem{Kamada:2010qe} 
  K.~Kamada, T.~Kobayashi, M.~Yamaguchi and J.~Yokoyama,
  %``Higgs G-inflation,''
  Phys.\ Rev.\ D {\bf 83}, 083515 (2011)
  [arXiv:1012.4238 [astro-ph.CO]].
%\bibitem{Kamada:2013bia} 
  K.~Kamada, T.~Kobayashi, T.~Kunimitsu, M.~Yamaguchi and J.~Yokoyama,
  %``Graceful exit from Higgs G-inflation,''
  Phys.\ Rev.\ D {\bf 88}, 123518 (2013)
  [arXiv:1309.7410 [hep-ph]].
\bibitem{Nakayama:2010kt} 
  K.~Nakayama and F.~Takahashi,
  %``Running Kinetic Inflation,''
  JCAP {\bf 1011}, 009 (2010)
  [arXiv:1008.2956 [hep-ph]].
\bibitem{Kamada:2012se} 
  K.~Kamada, T.~Kobayashi, T.~Takahashi, M.~Yamaguchi and J.~Yokoyama,
  %``Generalized Higgs inflation,''
  Phys.\ Rev.\ D {\bf 86}, 023504 (2012)
  [arXiv:1203.4059 [hep-ph]].
\bibitem{Mortonson:2014bja} 
  M.~J.~Mortonson and U.~Seljak,
  %``A joint analysis of Planck and BICEP2 B modes including dust polarization uncertainty,''
  JCAP10(2014)035
  [arXiv:1405.5857 [astro-ph.CO]].
R.~Flauger, J.~C.~Hill and D.~N.~Spergel,
  %``Toward an Understanding of Foreground Emission in the BICEP2 Region,''
  JCAP {\bf 1408}, 039 (2014)
  [arXiv:1405.7351 [astro-ph.CO]].
R.~Adam {\it et al.}  [Planck Collaboration],
  %``Planck intermediate results. XXX. The angular power spectrum of polarized dust emission at intermediate and high Galactic latitudes,''
  arXiv:1409.5738 [astro-ph.CO].
\bibitem{decigo}
N.\ Seto, S.\ Kawamura, and T.\ Nakamura, Phys.\ Rev.\ Lett. 
{\bf 87}, 221103 (2001);
S.~Kawamura {\it et al.},
  %``The Japanese space gravitational wave antenna DECIGO,''
  Class.\ Quant.\ Grav.\  {\bf 23}, S125 (2006).
  %%CITATION = CQGRD,23,S125;%%
\bibitem{Seto:2003kc}
  N.~Seto, J.~Yokoyama,
  %``Probing the equation of state of the early universe with a space laser interferometer,''
  J.\ Phys.\ Soc.\ Jap.\  {\bf 72}, 3082-3086 (2003).
  [gr-qc/0305096];
H.~Tashiro, T.~Chiba, M.~Sasaki,
  %``Reheating after quintessential inflation and gravitational waves,''
  Class.\ Quant.\ Grav.\  {\bf 21}, 1761-1772 (2004).
  [gr-qc/0307068];
 L.~A.~Boyle, P.~J.~Steinhardt,
  %``Probing the early universe with inflationary gravitational waves,''
  Phys.\ Rev.\  {\bf D77}, 063504 (2008).
  [astro-ph/0512014];
K.~Nakayama, S.~Saito, Y.~Suwa, J.~Yokoyama,
  %``Space laser interferometers can determine the thermal history of the early Universe,''
  Phys.\ Rev.\  {\bf D77}, 124001 (2008).
 [arXiv:0802.2452 [hep-ph]];
K.~Nakayama, S.~Saito, Y.~Suwa, J.~Yokoyama,
  %``Probing reheating temperature of the universe with gravitational wave background,''
  JCAP {\bf 0806}, 020 (2008).
  [arXiv:0804.1827 [astro-ph]];
K.~Nakayama, J.~Yokoyama,
  %``Gravitational Wave Background and Non-Gaussianity as a Probe of the Curvaton Scenario,''
  JCAP {\bf 1001}, 010 (2010).
  [arXiv:0910.0715 [astro-ph.CO]].
\bibitem{Kuroyanagi:2011fy} 
  S.~Kuroyanagi, K.~Nakayama and S.~Saito,
  %``Prospects for determination of thermal history after inflation with future gravitational wave detectors,''
  Phys.\ Rev.\ D {\bf 84}, 123513 (2011)
  [arXiv:1110.4169 [astro-ph.CO]].
\bibitem{chaoinf}
  A.D.\ Linde, Phys.\ Lett.\ B {\bf 129}, 177 (1983).
\bibitem{Murayama:1992ua}
  H.~Murayama, H.~Suzuki, T.~Yanagida and J.~Yokoyama,
  %``Chaotic inflation and baryogenesis by right-handed sneutrinos,''
  Phys.\ Rev.\ Lett.\  {\bf 70}, 1912 (1993);Phys.\ Rev.\  D {\bf 50}, 2356 (1994);  %%CITATION = PRLTA,70,1912;%%
M.~Kawasaki, M.~Yamaguchi and T.~Yanagida,
  %``Natural chaotic inflation in supergravity,''
  Phys.\ Rev.\ Lett.\  {\bf 85}, 3572 (2000);
%\bibitem{Kadota:2005mt} 
  K.~Kadota and J.~Yokoyama,
  %``D-term inflation and leptogenesis by right-handed sneutrino,''
  Phys.\ Rev.\ D {\bf 73}, 043507 (2006)
  [hep-ph/0512221].
\bibitem{Freese:1990rb} 
 K.~Freese, J.~A.~Frieman and A.~V.~Olinto,
  %``Natural inflation with pseudo - Nambu-Goldstone bosons,''
 Phys.\ Rev.\ Lett.\  {\bf 65}, 3233 (1990).
\bibitem{Kuroyanagi:2011iw} 
  S.~Kuroyanagi and T.~Takahashi,
  %``Higher Order Corrections to the Primordial Gravitational Wave Spectrum and its Impact on Parameter Estimates for Inflation,''
  JCAP {\bf 1110}, 006 (2011)
  [arXiv:1106.3437 [astro-ph.CO]].
  %%CITATION = ARXIV:1106.3437;%%
  %4 citations counted in INSPIRE as of 12 Mar 2014
\bibitem{Kuroyanagi:2008ye} 
  S.~Kuroyanagi, T.~Chiba and N.~Sugiyama,
  %``Precision calculations of the gravitational wave background spectrum from inflation,''
  Phys.\ Rev.\ D {\bf 79}, 103501 (2009)
  [arXiv:0804.3249 [astro-ph]];
  Phys.\ Rev.\ D {\bf 83}, 043514 (2011)
  [arXiv:1010.5246 [astro-ph.CO]].
\bibitem{Kuroyanagi:2014qaa} 
  S.~Kuroyanagi, S.~Tsujikawa, T.~Chiba and N.~Sugiyama,
  %``Implications of the B-mode Polarization Measurement for Direct Detection of Inflationary Gravitational Waves,''
  Phys.\ Rev.\ D {\bf 90}, 063513 (2014)
  [arXiv:1406.1369 [astro-ph.CO]].
\bibitem{Turner:1993vb}
  M.~S.~Turner, M.~J.~White, J.~E.~Lidsey,
  %``Tensor perturbations in inflationary models as a probe of cosmology,''
  Phys.\ Rev.\  {\bf D48}, 4613-4622 (1993).
  [astro-ph/9306029].
\bibitem{Nakayama:2008wy}
  K.~Nakayama, S.~Saito, Y.~Suwa, J.~Yokoyama,
  %``Probing reheating temperature of the universe with gravitational wave background,''
  JCAP {\bf 0806}, 020 (2008).
  [arXiv:0804.1827 [astro-ph]]. 
\bibitem{Jinno:2013xqa} 
  R.~Jinno, T.~Moroi and K.~Nakayama,
  %``Inflationary Gravitational Waves and the Evolution of the Early Universe,''
  JCAP {\bf 1401}, 040 (2014)
  [arXiv:1307.3010].
\bibitem{Kuroyanagi:2014nba} 
  S.~Kuroyanagi, T.~Takahashi and S.~Yokoyama,
  %``Blue-tilted Tensor Spectrum and Thermal History of the Universe,''
  arXiv:1407.4785 [astro-ph.CO].
\bibitem{Moroi:1994rs}
%\bibitem{Coughlan:1983ci} 
  G.~D.~Coughlan, W.~Fischler, E.~W.~Kolb, S.~Raby and G.~G.~Ross,
  %``Cosmological Problems for the Polonyi Potential,''
  Phys.\ Lett.\ B {\bf 131}, 59 (1983).
  T.~Moroi, M.~Yamaguchi and T.~Yanagida,
  %``On the solution to the Polonyi problem with 0 (10-TeV) gravitino mass in
  %supergravity,''
  Phys.\ Lett.\  B {\bf 342}, 105 (1995);
  %[arXiv:hep-ph/9409367];
  %%CITATION = PHLTA,B342,105;%% 
  %\cite{Kawasaki:1995cy}
%\bibitem{Kawasaki:1995cy}
  M.~Kawasaki, T.~Moroi and T.~Yanagida,
  %``Constraint on the Reheating Temperature from the Decay of the Polonyi
  %Field,''
  Phys.\ Lett.\  B {\bf 370}, 52 (1996);
  %[arXiv:hep-ph/9509399].
  %%CITATION = PHLTA,B370,52;%%
  %\cite{Nakamura:2007wr}
%\bibitem{Nakamura:2007wr}
  S.~Nakamura and M.~Yamaguchi,
  %``A Note on Polonyi Problem,''
  Phys.\ Lett.\  B {\bf 655}, 167 (2007).
  %[arXiv:0707.4538 [hep-ph]].
  %%CITATION = PHLTA,B655,167;%%  
\bibitem{Moroi:1999zb}
  T.~Moroi and L.~Randall,
  %``Wino cold dark matter from anomaly-mediated SUSY breaking,''
  Nucl.\ Phys.\  B {\bf 570}, 455 (2000);
  %[arXiv:hep-ph/9906527];
  %%CITATION = NUPHA,B570,455;%%
  %\cite{Kohri:2004qu}
  %\bibitem{Kohri:2004qu}
  K.~Kohri, M.~Yamaguchi and J.~Yokoyama,
  %``Production and dilution of gravitinos by modulus decay,''
  Phys.\ Rev.\  D {\bf 70}, 043522 (2004);
  %[arXiv:hep-ph/0403043].
  %%CITATION = PHRVA,D70,043522;%%
  %\cite{Kohri:2005ru}
%\bibitem{Kohri:2005ru}
  %K.~Kohri, M.~Yamaguchi and J.~Yokoyama,
  %``Neutralino dark matter from heavy gravitino decay,''
  Phys.\ Rev.\  D {\bf 72}, 083510 (2005);
  %[arXiv:hep-ph/0502211].
  %%CITATION = PHRVA,D72,083510;%%
%\bibitem{Endo:2006zj} 
  M.~Endo, K.~Hamaguchi and F.~Takahashi,
  %``Moduli-induced gravitino problem,''
  Phys.\ Rev.\ Lett.\  {\bf 96}, 211301 (2006)
  [hep-ph/0602061].
  %\cite{Nagai:2007ud}
  %\bibitem{Nagai:2007ud}
  M.~Nagai and K.~Nakayama,
  %``Nonthermal dark matter in mirage mediation,''
  Phys.\ Rev.\  D {\bf 76}, 123501 (2007).
  %[arXiv:0709.3918 [hep-ph]].
  %%CITATION = PHRVA,D76,123501;%%
\bibitem{Seto:2005qy} 
  N.~Seto,
  %``Correlation analysis of stochastic gravitational wave background around 0.1-1 Hz,''
  Phys.\ Rev.\ D {\bf 73}, 063001 (2006)
  [gr-qc/0510067].
\bibitem{Fisherref} 
  S.~Kuroyanagi, C.~Gordon, J.~Silk and N.~Sugiyama,
  %``Forecast Constraints on Inflation from Combined CMB and Gravitational Wave Direct Detection Experiments,''
  Phys.\ Rev.\ D {\bf 81}, 083524 (2010)
  [Erratum-ibid.\ D {\bf 82}, 069901 (2010)]
  [arXiv:0912.3683 [astro-ph.CO]].
%\bibitem{Kuroyanagi:2013ns} 
  S.~Kuroyanagi, C.~Ringeval and T.~Takahashi,
  %``Early Universe Tomography with CMB and Gravitational Waves,''
  Phys.\ Rev.\ D {\bf 87}, 083502 (2013)
  [arXiv:1301.1778 [astro-ph.CO]].
\bibitem{Jinno:2014qka} 
  R.~Jinno, T.~Moroi and T.~Takahashi,
  %``Studying Inflation with Future Space-Based Gravitational Wave Detectors,''
  arXiv:1406.1666 [astro-ph.CO].
\bibitem{Allen:1997ad} 
  B.~Allen and J.~D.~Romano,
  %``Detecting a stochastic background of gravitational radiation: Signal processing strategies and sensitivities,''
  Phys.\ Rev.\ D {\bf 59}, 102001 (1999)
  [gr-qc/9710117].
\bibitem{Farmer:2003pa}
  A.~J.~Farmer and E.~S.~Phinney,
  %``The Gravitational Wave Background from Cosmological Compact Binaries,''
  Mon.\ Not.\ Roy.\ Astron.\ Soc.\  {\bf 346}, 1197 (2003)
  [arXiv:astro-ph/0304393].
%\cite{Maggiore:1999vm}
\bibitem{Maggiore:1999vm} 
  M.~Maggiore,
  %``Gravitational wave experiments and early universe cosmology,''
  Phys.\ Rept.\  {\bf 331}, 283 (2000)
  [gr-qc/9909001].
  %%CITATION = GR-QC/9909001;%%
\bibitem{maggiore}
 M.\ Maggiore, ``Gravitational Waves Volume 1: Theory and Experiments,''
Oxford University Press (Oxford, 2008)
\bibitem{Kawamura:2011zz} 
  S.~Kawamura, M.~Ando, N.~Seto, S.~Sato, T.~Nakamura, K.~Tsubono, N.~Kanda and T.~Tanaka {\it et al.},
  %``The Japanese space gravitational wave antenna: DECIGO,''
  Class.\ Quant.\ Grav.\  {\bf 28}, 094011 (2011).
  %%CITATION = CQGRD,28,094011;%%
%\cite{Yagi:2011wg}
\bibitem{Yagi:2011wg} 
  K.~Yagi and N.~Seto,
  %``Detector configuration of DECIGO/BBO and identification of cosmological neutron-star binaries,''
  Phys.\ Rev.\ D {\bf 83}, 044011 (2011)
  [arXiv:1101.3940 [astro-ph.CO]].
  %%CITATION = ARXIV:1101.3940;%%
%\bibitem{3bai}  Amardi 10 talk?
\bibitem{Lyth} 
D.~H.~Lyth,
  %``What would we learn by detecting a gravitational wave signal in the cosmic microwave background anisotropy?,''
  Phys.\ Rev.\ Lett.\  {\bf 78}, 1861 (1997)
  [hep-ph/9606387].
    %\cite{Kawasaki:2000yn}
\bibitem{Kawasaki:2000yn} 
  M.~Kawasaki, M.~Yamaguchi and T.~Yanagida,
  %``Natural chaotic inflation in supergravity,''
  Phys.\ Rev.\ Lett.\  {\bf 85}, 3572 (2000)
  [hep-ph/0004243].
  %%CITATION = HEP-PH/0004243;%%
  %\cite{Nakayama:2013txa}
\bibitem{Nakayama:2013txa} 
  K.~Nakayama, F.~Takahashi and T.~T.~Yanagida,
  %``Polynomial Chaotic Inflation in Supergravity,''
  JCAP {\bf 1308}, 038 (2013)
  [arXiv:1305.5099 [hep-ph]].
  %%CITATION = ARXIV:1305.5099,;%%
\bibitem{Bolz:2000fu} 
  M.~Bolz, A.~Brandenburg and W.~Buchmuller,
  %``Thermal production of gravitinos,''
  Nucl.\ Phys.\ B {\bf 606}, 518 (2001)
  [Erratum-ibid.\ B {\bf 790}, 336 (2008)]
  [hep-ph/0012052];
  %%CITATION = HEP-PH/0012052;%%
%\cite{Pradler:2006qh}
%\bibitem{Pradler:2006qh} 
  J.~Pradler and F.~D.~Steffen,
  %``Thermal gravitino production and collider tests of leptogenesis,''
  Phys.\ Rev.\ D {\bf 75}, 023509 (2007)
  [hep-ph/0608344];
  %%CITATION = HEP-PH/0608344;%%
%\cite{Pradler:2006hh}
%\bibitem{Pradler:2006hh} 
  %J.~Pradler and F.~D.~Steffen,
  %``Constraints on the Reheating Temperature in Gravitino Dark Matter Scenarios,''
  Phys.\ Lett.\ B {\bf 648}, 224 (2007)
  [hep-ph/0612291];
  %%CITATION = HEP-PH/0612291;%%
  %\cite{Rychkov:2007uq}
%\bibitem{Rychkov:2007uq} 
  V.~S.~Rychkov and A.~Strumia,
  %``Thermal production of gravitinos,''
  Phys.\ Rev.\ D {\bf 75}, 075011 (2007)
  [hep-ph/0701104].
  %%CITATION = HEP-PH/0701104;%%
  
\bibitem{Kawasaki:2004qu}
  M.~Kawasaki, K.~Kohri and T.~Moroi,
  %``Hadronic decay of late-decaying particles and big-bang nucleosynthesis,''
  Phys.\ Lett.\  B {\bf 625}, 7 (2005);
  %[arXiv:astro-ph/0402490].
  %%CITATION = PHLTA,B625,7;%%
  %M.~Kawasaki, K.~Kohri and T.~Moroi,
  %``Big-bang nucleosynthesis and hadronic decay of long-lived massive
  %particles,''
  Phys.\ Rev.\  D {\bf 71}, 083502 (2005).
   %\cite{Bolz:2000fu}
  \bibitem{Moroi:1993mb}
  T.~Moroi, H.~Murayama and M.~Yamaguchi,
  %``Cosmological constraints on the light stable gravitino,''
  Phys.\ Lett.\  B {\bf 303}, 289 (1993);
  %%CITATION = PHLTA,B303,289;%%
A.~de Gouvea, T.~Moroi and H.~Murayama,
 %``Cosmology of supersymmetric models with low-energy gauge mediation,''
  Phys.\ Rev.\  D {\bf 56}, 1281 (1997).
% [arXiv:hep-ph/9701244].
  %%CITATION = PHRVA,D56,1281;%%

%\cite{Ibe:2011aa}
\bibitem{Ibe:2011aa} 
  M.~Ibe and T.~T.~Yanagida,
  %``The Lightest Higgs Boson Mass in Pure Gravity Mediation Model,''
  Phys.\ Lett.\ B {\bf 709}, 374 (2012)
  [arXiv:1112.2462 [hep-ph]].
  %%CITATION = ARXIV:1112.2462;%%
  %\cite{ArkaniHamed:2012gw}
\bibitem{ArkaniHamed:2012gw} 
  N.~Arkani-Hamed, A.~Gupta, D.~E.~Kaplan, N.~Weiner and T.~Zorawski,
  %``Simply Unnatural Supersymmetry,''
  arXiv:1212.6971 [hep-ph].
  %%CITATION = ARXIV:1212.6971;%%
  %66 citations counted in INSPIRE as of 09 Apr 2014
%\cite{Arvanitaki:2012ps}
\bibitem{Arvanitaki:2012ps} 
  A.~Arvanitaki, N.~Craig, S.~Dimopoulos and G.~Villadoro,
  %``Mini-Split,''
  JHEP {\bf 1302}, 126 (2013)
  [arXiv:1210.0555 [hep-ph]].
  %%CITATION = ARXIV:1210.0555;%%

  %\cite{Nakayama:2014xca}
\bibitem{Nakayama:2014xca} 
  K.~Nakayama, F.~Takahashi and T.~T.~Yanagida,
  %``Gravitino Problem in Supergravity Chaotic Inflation and SUSY Breaking Scale after BICEP2,''
  Phys.\ Lett.\ B {\bf 734}, 358 (2014)
  [arXiv:1404.2472 [hep-ph]].
  %%CITATION = ARXIV:1404.2472;%%
  

\end{thebibliography}
\end{document}